\documentclass{article}

\usepackage{arxiv}

\usepackage[utf8]{inputenc} % allow utf-8 input
\usepackage[T1]{fontenc}    % use 8-bit T1 fonts
\usepackage{hyperref}       % hyperlinks
\usepackage{url}            % simple URL typesetting
\usepackage{booktabs}       % professional-quality tables
\usepackage{amsfonts}       % blackboard math symbols
\usepackage{nicefrac}       % compact symbols for 1/2, etc.
\usepackage{microtype}      % microtypography
\usepackage{lipsum}		% Can be removed after putting your text content
\usepackage{graphicx}
\usepackage{natbib}
\usepackage{doi}

\title{Beyond Rules:\\Towards Basso Continuo Personal Style Identification}

\author{ \href{https://orcid.org/0009-0001-3201-9983}{\includegraphics[scale=0.06]{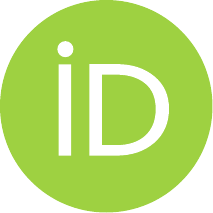}\hspace{1mm}Adam~Štefunko} \\
	Faculty of Mathematics and Physics\\
	Charles University\\
	Prague, Czech Republic \\
	\texttt{stefunko@ufal.mff.cuni.cz} \\
    \And
	\href{https://orcid.org/0000-0002-9207-567X}{\includegraphics[scale=0.06]{orcid.pdf}\hspace{1mm}Jan~Hajič~jr.} \\
	Faculty of Mathematics and Physics\\
	Charles University\\
	Prague, Czech Republic \\
	\texttt{hajicj@ufal.mff.cuni.cz} \\
}

\renewcommand{\shorttitle}{Beyond Rules}

\hypersetup{
pdftitle={Beyond Rules: Towards Basso Continuo Personal Style Identification},
pdfsubject={cs.SD},
pdfauthor={Adam Štefunko, Jan Hajič jr.},
pdfkeywords={Basso continuo, Historically Informed Practice, Improvisation, Performance analysis},
}

\begin{document}
\maketitle

\begin{abstract}
A central part of the contemporary Historically Informed Practice movement is \textit{basso continuo}, an improvised accompaniment genre with its traditions originating in the baroque era and actively practiced by many keyboard players nowadays. Although computational musicology has studied the theoretical foundations of \textit{basso continuo} expressed by harmonic and voice-leading rules and constraints, characteristics of \textit{basso continuo} as an active performing art have been largely overlooked mostly due to a lack of suitable performance data that could be empirically analyzed. This has changed with the introduction of The Aligned Continuo Realization Dataset (ACoRD) and the \textit{basso continuo} realization-to-score alignment. \textit{Basso continuo} playing is shaped by stylistic traditions coming from historical treatises, but it also may provide space for showcasing individual performance styles of its practitioners. In this paper, we attempt to explore the question of the presence of personal styles in the \textit{basso continuo} realizations of players in the ACoRD dataset. We use a historically informed structured representation of \textit{basso continuo} performance pitch content called \textit{griffs} and Support Vector Machines to see whether it is possible to classify players based on their performances. The results show that we can identify players from their performances. In addition to the player classification problem, we discuss the elements that make up the individual styles of the players. 
\end{abstract}

% keywords can be removed
%basso continuo, historically informed performance, improvisation, performance analysis

\section{Introduction} 
Harmonic thinking of baroque music has been largely shaped by \textit{basso continuo}, which not only served as an improvisatory accompaniment style, but it influenced how several generations of composers understood harmony \cite{holtmeier2007heinichen}.
The practice of \textit{basso continuo} consists of improvising multiple musical parts above a given bass part, following learned rules and typical stylistic patterns. This process is called \textit{basso continuo realization}. Based on historical treatises, we can trace three distinctly prominent historic \textit{continuo} realization styles: French \cite{dandrieu1719principes, de1710nouveau}, German \cite{heinichen1728general, mattheson1731grosse} and Italian \cite{gasparini1708armonico, geminianiTheArt}.
These styles are characterized by different elements of texture like typical chord positions and transitions, number of parts, added ornamentation and occasional melodic elements \cite{christensen_18th_2002}.

But \textit{basso continuo} practice is 
not only a mere historical subject: it is
still a living tradition, studied and practiced by many historical keyboard and plucked instrument players within the Historically Informed Performance movement. 
Given the reliance on textual sources that emphasize how to conform to these harmonic and stylistic expectation, \textit{continuo} realization is sometimes thought of just a procedure of following harmonic and contrapuntal rules. However, according to historical treatises, exercise and imitation of other players' styles is equally important. Although \textit{continuo} improvisation is harmonically and stylistically constrained, it still provides some freedom to the player. 

Whereas the historical styles of \textit{basso continuo} are preserved in a written form, there has not been much data of keyboard players performing \textit{continuo} realizations in our era. This has changed with the first collection of \textit{basso continuo} performances recorded in a symbolic format (MIDI) -- The Aligned Continuo Realization Dataset (ACoRD)\footnote{https://hdl.handle.net/11234/1-5963}. We are now able to study the realizations of players today, and we can start examining the extent of individual creative agency within the framework given by what \textit{continuo} requires \cite{mortensen1996unerringly}. 

In this paper, we deal with the question of individuality in \textit{basso continuo} realizations. We attempt to classify realizations by player to ultimately answer the following questions: Is there enough space within the constraints of \textit{basso continuo} for distinctively personal realization choices? Can we empirically trace imprints of individuality in \textit{continuo} realizations of different players?

A previous research \cite{MEC} suggests that \textit{griffs}, a structured representation of a \textit{continuo} realization (Figure~\ref{fig:griff-extraction}), has the potential of showing whether players are consistent within their own \textit{continuo} performances in comparison to performances of two different players. In this paper, we build upon this result and investigate whether it is possible to use performance \textit{griff} profiles to trace players' individuality by classifying them using \textit{Support Vector Machines} (SVMs). We then further analyze what elements make identifiable personal style on the \textit{griff} level.

\section{Related Work}

The few previous works on \textit{continuo} in digital musicology have mostly focused on \textit{basso continuo} as a theoretical concept. This includes \textit{basso continuo} generation works using dynamic programming \cite{niitsuma2011development} or decision trees \cite{wead2007computer}, and machine learning-driven automated figured bass annotation \cite{ju2020automatic}. Machine learning methods also lie within the larger scope of general music generation, where we may observe a few attempts of stylistically-rooted music generation using different architectures (recurrent neural networks \cite{hadjeres2017deepbach, liang2017automatic, marinescu2019bach}, transformers \cite{yao_2025_generality}, large language models \cite{zhu_GPT}). Probably the most famous example -- a convolutional network model \textit{Coconet} \cite{huang2017counterpoint} -- focused on generating chorales in the style of J. S. Bach. However, none of these systems deal with live musical performances. 

\section{Materials and Methods}
\subsection{ACoRD Dataset}
The dataset contains 175 MIDI recordings. These recordings are realizations of 5 \textit{continuo} lines written in musical scores, each played 5 times by each of 7 harpsichordists (professionals and students). The overall number of notes in a \textit{continuo} line spans from 82 to 135, with the average being approximately 105 \cite{stefunko_2025_AIMC}. An automatic dynamic time warping-based system \cite{stefunko_2025_AIMC} provides alignments of performances in the dataset to their respective \textit{continuo} lines. 
The alignments provided by the automatic system are not perfect \cite{chiruthapudi_2025_challenges}, but they do provide a sufficiently precise way of grouping performance notes that belong to the same score note, and aligning the bass notes --- which we then rely on for feature extraction --- is nearly perfect.

\subsection{Performance Representation: \textit{Griffs}}
\begin{figure}[h]
  \centering
  \includegraphics[width=0.8\linewidth]{figures/Griff_formation_2.pdf}
  \caption{\textit{Griffs} are extracted from aligned performances (a). A \textit{griff} is a sequence of vectors formed by notes appearing together within a time window (b), transformed from pitches to intervals (c), and encoded as strings (d) that can form \textit{n}-grams (e).}
  \label{fig:griff-extraction}
\end{figure}

\textit{Griffs} represent temporal structure of notes aligned with the same \textit{basso continuo} score note without retaining their absolute onset timings.
Notes are ordered according to their onset times. Notes that appear within a certain window\footnote{Window size is a parameter that can be set to different values. We use 35\textit{ms}, a value suggested by the original paper on the \textit{griff} representation \cite{MEC}.} are put together to a single vector, notes that come afterwards -- within the next window -- form the next vector. \textit{Griff} is a sequence of these vectors. Pitches in \textit{griffs} are turned to intervals from the bass note they are aligned to. This conversion enables working with the \textit{griff} shapes regardless of their transposition and see their similarity across all the possible tonalities. As the last step, \textit{griffs} are encoded as strings by joining the intervals that form a single vector by an underscore and subsequent vectors by a vertical line (see Figure~\ref{fig:griff-extraction}). \textit{Griffs} that have the same content, and therefore encoded as the same string, form a \textit{griff} type.
\textit{Griffs} encompass the tactile properties of \textit{basso continuo} playing -- actually pressing several keys at a keyboard using learned patterns. This follows the assumption that keyboard \textit{continuo} playing does not necessarily follow contrapuntal rules, but several treatises show typical chord patterns, including works of Georg Muffat \cite{muffat1699regulae}, Jean-François Dandrieu \cite{dandrieu1719principes} and Francesco Geminiani \cite{geminianiTheArt}. 
Furthermore, \textit{griff} \textit{n}-grams can be formed by concatenating \textit{n} subsequent \textit{griff}-encoding strings using a different symbol, which in our case is a hashtag. 
In the opposite direction, we can get rid of the structural aspect of \textit{griffs} and simply convert the pitch of every performance note to an interval from the score note to which this performance note is aligned to, to form the \textit{interval} representation.

\begin{table}[]
\begin{tabular}{@{}lcccc@{}}
\toprule
\textbf{Vocab. sizes}     & Intervals & \textit{Griffs} & \textit{Griff} 2-grams & \textit{Griff} 3-grams \\ \midrule
Score 001     & 48        & 1200            & 2300                   & 2607                    \\
Score 002     & 54        & 2231            & 3356                   & 3640                    \\
Score 003     & 46        & 911             & 1903                   & 2311                    \\
Score 004     & 50        & 2125            & 3157                   & 3317                    \\
Score 005     & 45        & 2223            & 3761                   & 4150                    \\ \midrule
Whole Dataset & 54        & 7061            & 14107                  & 15989                   \\ \bottomrule
\end{tabular}
\caption{Vocabulary size for individual scores and the whole dataset based on the representation used. The total number of \textit{griff} tokens is 17152 after filtering empty \textit{griffs}.} 
\label{tab:vocabulary-sizes}
\end{table}

\subsection{Player Classification with SVMs}
For player classification, we first obtain a \textit{griff} profile of each performance by counting how many times was each \textit{griff} type used in the performance. 
Since \textit{griffs} are extracted from performances aligned to score notes, each performance of the same score should contain the same number of \textit{griffs}\footnote{This includes empty \textit{griffs} for note deletions and \textit{griffs} only containing the bass note. Even though they are the most populated \textit{griff} types, they are excluded from the \textit{griff} profiles. Although they can be relevant for some tasks, they contain no pitch content information, which we primarily focus on.}.     
These distributions of \textit{griffs} are then turned into \textit{bag of words} (BOW) representations of each performance above the common vocabulary of all \textit{griffs} present in the performances. 
This forms a feature space in which we solve the player classification problem using SVMs. We equivalently turn \textit{griff} bigrams, trigrams and intervals into BOW matrices.
Vocabulary sizes for different representations are shown in Table~\ref{tab:vocabulary-sizes}.
We use the \texttt{SVC} classifier from \texttt{sklearn.svm} \cite{scikit-learn}, trained  with linear, polynomial, RBF and sigmoid kernels on player-labeled performances.

We evaluate the classification experiments using 5-fold stratified cross-validation. We also conduct player-focused classification experiments in which we want to see how accurately we are able to classify performances of a selected player.
In each independent run of these experiments, we take away one performance of the selected player for testing and train on all the other performances in the relevant part of the dataset.
We, then, count in how many runs was a test performance correctly classified. 
High accuracy scores mean that there are distinct personal styles captured by the given representation. Low accuracy scores can be interpreted in two different ways: 
either all players play similarly or performances of a given player are too divergent.   

\begin{table}[]
 \begin{tabular}{@{}lcccc@{}}
\toprule
 & Intervals & \textit{Griffs} & \textit{Griff} Bigrams & \textit{Griff} Trigrams \\ \midrule
Score 001                    & 0.83      & 0.91   & 0.97          & 0.63           \\
Score 002                    & 0.94      & 0.97   & 0.94          & 0.63           \\
Score 003                    & 0.74      & 0.94   & 0.94          & 0.69           \\
Score 004                    & 0.91      & 0.97   & 0.71          & 0.31           \\
Score 005                    & 0.94      & 0.97   & 0.89          & 0.71           \\ \midrule
Whole Dataset                & 0.60      & 0.87   & 0.73          & 0.49           \\ \bottomrule
\end{tabular}
 \caption{Stratified 5-fold cross-validation accuracy of Linear SVM player classification on each of the scores and on the whole dataset using different performance representations (intervals, \textit{griffs}, \textit{griff} bigrams and trigrams)}
 \label{tab:player_class}
\end{table}

\section{Player Classification Results}
We measured player classification accuracy on performances of each score separately and on the whole dataset. Accuracies in Table~\ref{tab:player_class} show that \textit{griffs} are the best performing representation not only on the whole dataset, but also on four out of five scores. We report results only for the linear kernel, as other kernels show the same or worse classification accuracies. Player-focused experiments on scores using \textit{griffs} show high accuracies (0.8 -- 1.0) for for all but player \texttt{8780} (mean on scores 0.68).

\begin{figure}[h]
  \centering
  \includegraphics[width=0.8\linewidth]{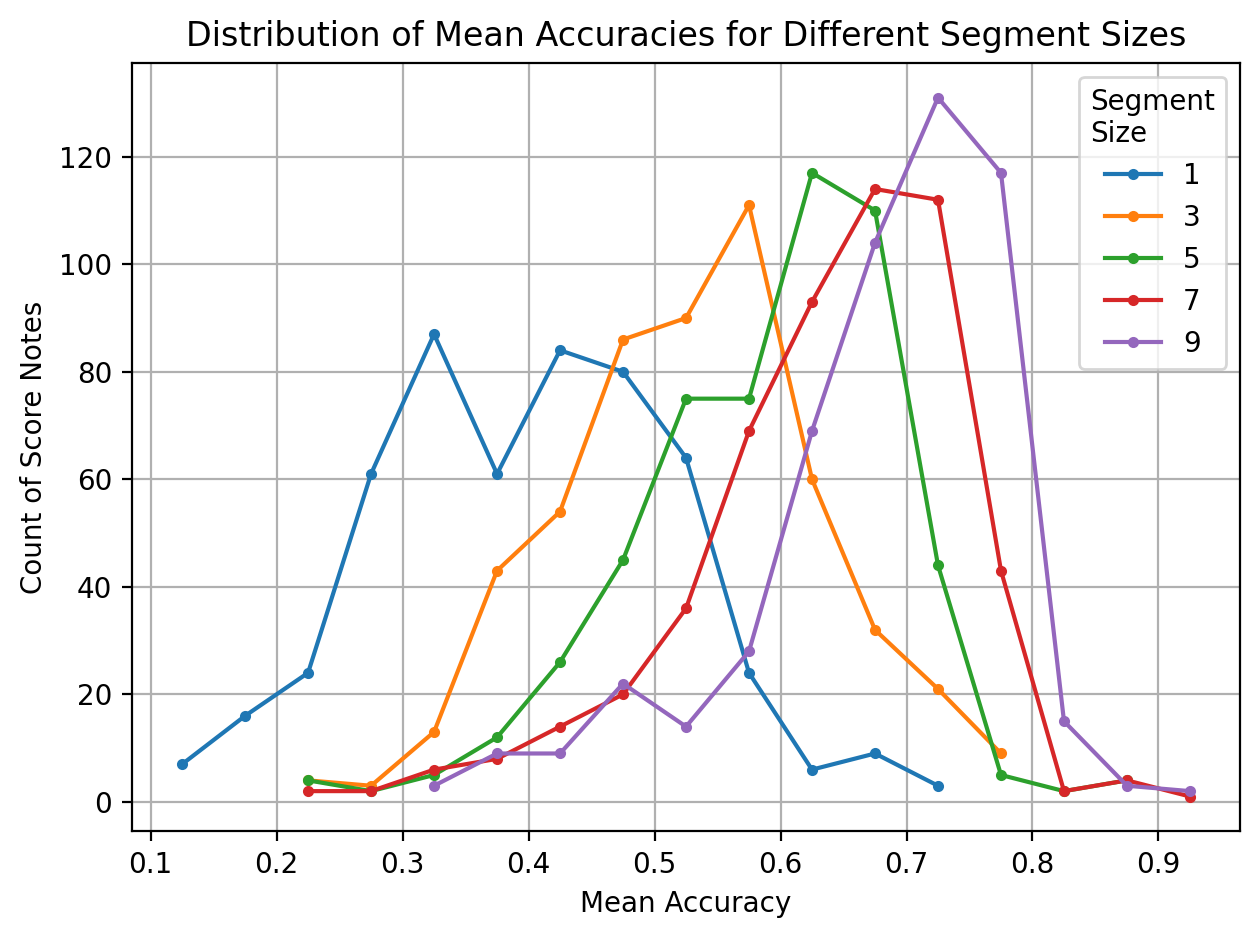}
  \caption{Accuracy distributions of different score notes using segments of different size aggregated over the whole dataset.}
  \label{fig:accuracy-histogram}
\end{figure}
\begin{figure*}[t]
  \centering
  \includegraphics[width=\linewidth]{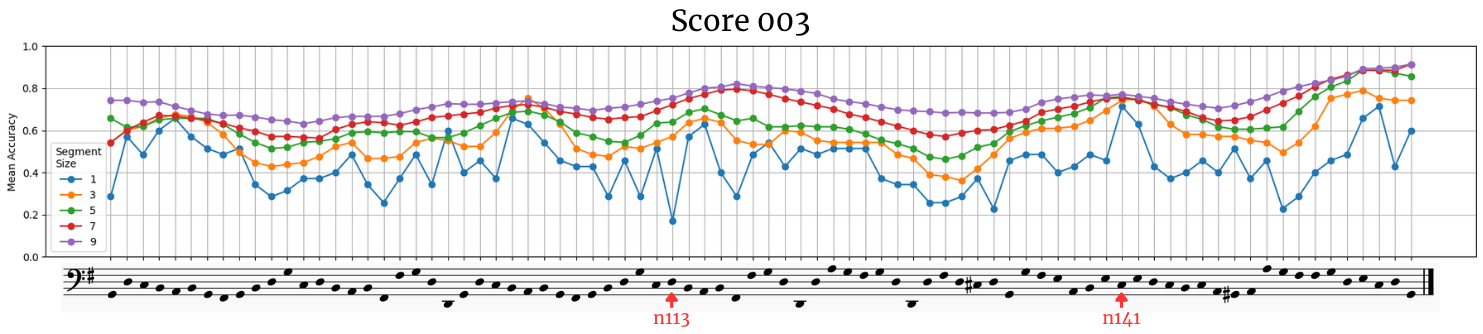}
  \caption{Score note mean accuracies using segments of different lengths. Musical staves below each plot represent score note pitches. Notes marked with red arrows and note IDs are used for a case study in Section~\ref{sec:elements}.}
  \label{fig:note_id_accur}
\end{figure*}

\section{Exploring Elements of Players' Styles}
\label{sec:elements}

Can we trace some player-specific elements specific in the \textit{griff} content? 
The easiest and clearest approach is to analyze style in terms of \textit{griff} choices uniquely but consistently applied by individual players in specific musical situations. 
To see how this analytical approach of ``personal \textit{griff} vocabularies'' can be applied, we must find out at what resolution can we localize individuality in the score.
If the overall success of player classification relies on such specific locations in the scores rather than a more nebulous persistent style, we should be able to classify players based on their personal preferred \textit{griffs} in such locations alone.
In this experiment, we put together all performances of the same segment and train SVM classifier exclusively on them. 
Figure~\ref{fig:accuracy-histogram} shows that the distributions of accuracies for different segment sizes roughly follow skew normal distribution.
Figure~\ref{fig:note_id_accur} shows the course of the mean classification accuracies for Score 003.

Rather than any localised spikes in accuracy, the most stand-out features seem to rather be regions of \textit{lower} accuracy. 
Observing a generally lower accuracy provides us with some space to look at \textit{griff} distributions in a few particular cases where imprints of personal styles do not seem to be so present and compare them with higher accuracy segments.
Figure~\ref{fig:note_id_accur} reveals score notes whose \textit{griff} contents are worth comparing. 

Table~\ref{tab:selected_note_accuracies} compares selected score notes and how their accuracies relate to the number of \textit{griff} types and how many times on average is each \textit{griff} type used to realize them. Score note \texttt{n238} from \texttt{Score 004}\footnote{ Some note IDs are higher than the actual number of notes in the \textit{basso continuo} line. This is because the scores in the ACoRD dataset contain and index full scores including other vocal or instrumental parts.} is apparently an extreme case in which a different griff is used in almost every performance. Note \texttt{n39} from \texttt{Score 002} also has a rather large number of \textit{griff} types played above it. 

\begin{table}[]
    \begin{tabular}{@{}lllll@{}}
    \toprule
    \begin{tabular}[c]{@{}l@{}}Score Name:\\ Note ID\end{tabular} & \begin{tabular}[c]{@{}l@{}}Note\\ Spelling\end{tabular} & \begin{tabular}[c]{@{}l@{}}Classification \\ Accuracy\end{tabular} & \begin{tabular}[c]{@{}l@{}}Number of \\ Griff Types\end{tabular} & \begin{tabular}[c]{@{}l@{}}Mean Griff \\ Type Usage\end{tabular} \\ \midrule
    
    003: n113                                                     & D3                                                      & 0.17                                                               & 11                                                               & 3.18                                                             \\
    
    004: n238                                                     & D3                                                      & 0.20                                                               & 34                                                               & 1.03 
    \\
    
    002: n39                                                      & F\#\#3                                                  & 0.31                                                               & 23                                                               & 1.52                                                             \\
    \midrule
    003: n141                                                     & C3                                                      & 0.71                                                               & 16                                                               & 2.19                                                  
    \\ \bottomrule
    \end{tabular}
    \caption{Selected notes with low mean player classification accuracy (including a note with a higher mean accuracy for comparison) with a number of different \textit{griff} types associated with them and the mean of how many times each \textit{griff} type is played. 
    }
    \label{tab:selected_note_accuracies}
\end{table}

\begin{figure}[h]
  \centering
  \includegraphics[width=0.8\linewidth]{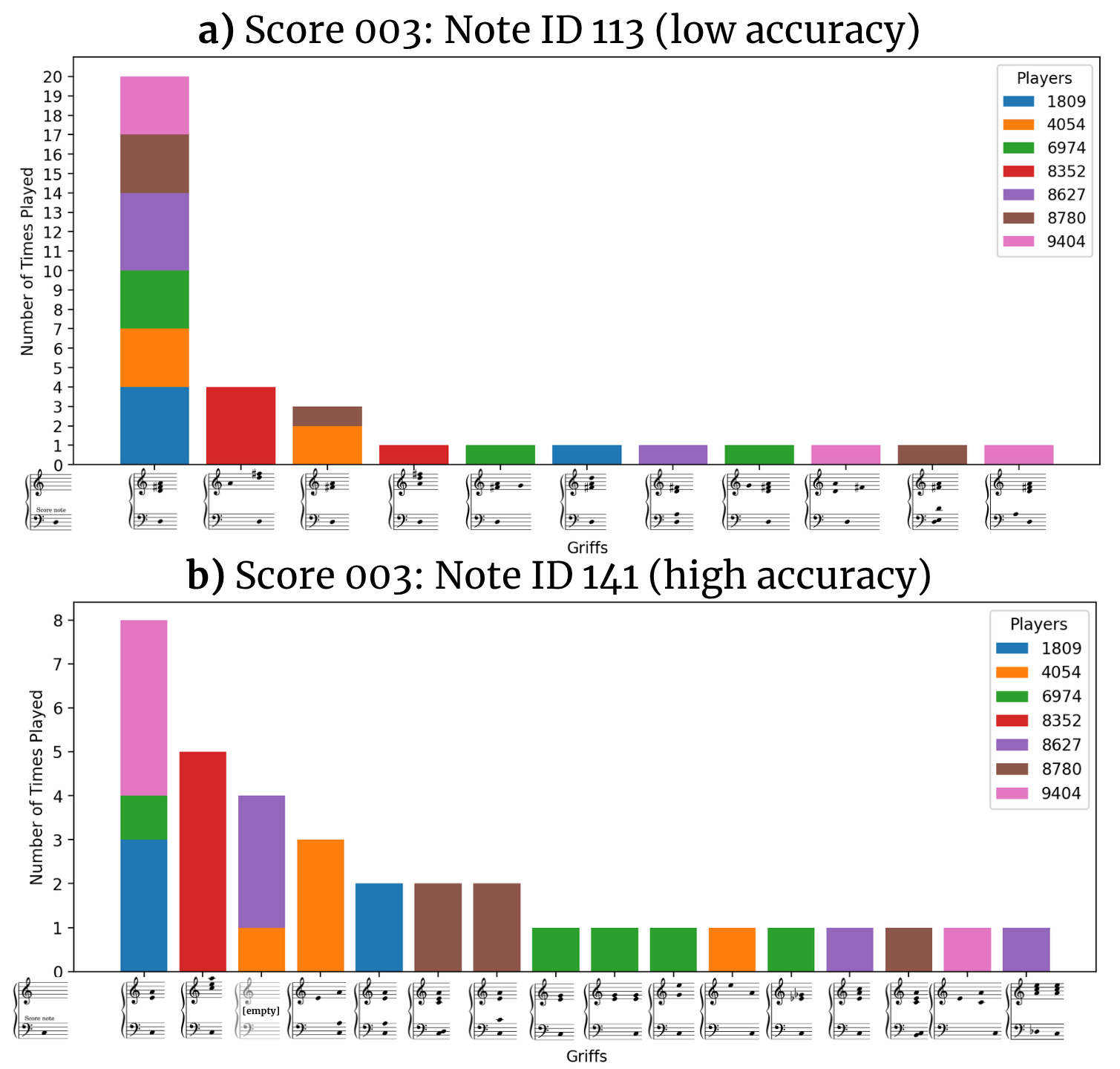}
  \caption{\textit{Griff} type distributions among different players for a high and a low accuracy note of Score 003. Each bar on the \textit{x}-axis represents a different \textit{griff} type. 
  }
  \label{fig:case-study-distributions}
\end{figure}

However, note \texttt{n113} from \texttt{Score 003} is not so clear. Given that the number of times a \textit{griff} type is played is on average slightly more than 3, this note would have a chance for higher accuracy if a lot of these \textit{griffs} were distributed among players with only little intersection. Therefore, it seems that the distribution of \textit{griffs} among players is responsible for the low mean accuracy. This is visible especially in comparison with a high accuracy note \texttt{n141} from the same score (Figure~\ref{fig:case-study-distributions}). Although both notes have many \textit{griffs} that occur only once, the difference between these score notes is that most performances have realized the score note \texttt{n113} using one \textit{griff} type, which occurs in 20 performances and is played by all but one player. Score note \texttt{n141} has, on the other hand, a more even distribution.

\section{Conclusion}
\textit{Basso continuo}, as an improvisatory practice, is rooted in historical stylistic traditions which provide a constrained framework in which \textit{basso continuo} realizations are improvised. However, \textit{basso continuo} also gives its practitioner some freedom and the possibility of developing a personal style distinct from other players.
In this paper, we studied the extent of individuality within this constrained environment, showing that it is possible to tell individual players apart by their realizations, using the machine learning classification method SVM and the \textit{griff} representation of the performances.
We further attempted to localize the players' individuality within the performances. Classification of differently sized segments revealed that \textit{griff} type diversity and distribution are important elements of player classification.
%These are some of the first steps on the path toward u
Understanding how modern-day players perform \textit{basso continuo} from the empirical perspective is of a particular importance since there are very few ways to record how this improvisatory art is practiced. A future research may provide insights valuable for new \textit{basso continuo} practitioners, and serve as a starting point toward analysis of other improvisatory genres.

\bibliographystyle{unsrtnat}
\bibliography{paper_bib}  %%% Uncomment this line and comment out the ``thebibliography'' section below to use the external .bib file (using bibtex) .

@inproceedings{MEC,
  author    =   {Štefunko, Adam and
                Cancino-Chacón, Carlos Eduardo and
                Hajič jr., Jan},
  title     = {On Every Note a Griff: Looking for a Useful Representation of Basso Continuo Performance Style},
  booktitle = {Music Encoding Conference 2026},
  year      = {2026},
venue        = {Tokio, Japan},
}

@inproceedings{chiruthapudi_2025_challenges,
  author       = {Chiruthapudi, Suhit and
                  Štefunko, Adam and
                  Hajič jr., Jan and
                  Cancino-Chacón, Carlos Eduardo},
  title        = {Challenges in Basso Continuo Performance-to-Score
                   Alignment
                  },
  booktitle    = {Proceedings of the 17th International Symposium on
                   Computer Music Multidisciplinary Research
                  },
  year         = 2025,
  pages        = {924-936},
  publisher    = {Zenodo},
  month        = nov,
  venue        = {London, United Kingdom},
  doi          = {10.5281/zenodo.17502511},
  url          = {https://doi.org/10.5281/zenodo.17502511},
}

@book{christensen_18th_2002,
	address = {Kassel},
	title = {18th {Century} {Continuo} {Playing}: {A} {Historical} {Guide} to the {Basics}},
	isbn = {979-0006505494},
	language = {English},
	publisher = {Bärenreiter},
	author = {Christensen, Jesper Bøje},
	year = {2002},
}

@book{dandrieu1719principes,
  title={Principes de l'accompagnement du clavecin},
  author={Dandrieu, Jean Fran{\c{c}}ois},
  year={1718},
  publisher={Paris: M. Bayard}
}

@book{de1710nouveau,
  title={Nouveau trait{\'e} de l'accompagnement du clavecin, de l'orgue, et des autres instruments},
  author={de Saint-Lambert, Michel},
  year={1710},
  publisher={E. Roger}
}

@book{gasparini1708armonico,
  title={L'armonico pratico al cimbalo},
  author={Gasparini, Francesco},
  year={1708},
  publisher={Venezia: Antonio Bortoli}
}

@book{geminianiTheArt,
  title={{The Art of Accompaniment, Op.11}},
  author={Geminiani, Francesco},
  year={n.d.},
  publisher={London: John Johnson}
}

@article{holtmeier2007heinichen,
  title={Heinichen, Rameau, and the Italian thoroughbass tradition: Concepts of tonality and chord in the rule of the octave},
  author={Holtmeier, Ludwig},
  journal={Journal of Music Theory},
  volume={51},
  number={1},
  pages={5--49},
  year={2007},
  publisher={Duke University Press}
}

@inproceedings{hadjeres2017deepbach,
  title={Deepbach: a steerable model for bach chorales generation},
  author={Hadjeres, Ga{\"e}tan and Pachet, Fran{\c{c}}ois and Nielsen, Frank},
  booktitle={International conference on machine learning},
  pages={1362--1371},
  year={2017},
  organization={PMLR}
}

@book{heinichen1728general,
  title={Der General-Bass in der Composition},
  author={Heinichen, Johann David},
  year={1728},
  publisher={Dresden: the author}
}

@inproceedings{huang2017counterpoint,
  author       = {Cheng{-}Zhi Anna Huang and
                  Tim Cooijmans and
                  Adam Roberts and
                  Aaron C. Courville and
                  Douglas Eck},
  editor       = {Sally Jo Cunningham and
                  Zhiyao Duan and
                  Xiao Hu and
                  Douglas Turnbull},
  title        = {Counterpoint by Convolution},
  booktitle    = {Proceedings of the 18th International Society for Music Information
                  Retrieval Conference, {ISMIR} 2017, Suzhou, China, October 23-27,
                  2017},
  pages        = {211--218},
  year         = {2017},
  url          = {https://ismir2017.smcnus.org/wp-content/uploads/2017/10/187\_Paper.pdf},
  bibsource    = {dblp computer science bibliography, https://dblp.org}
}

@inproceedings{ju2020automatic,
  title={Automatic figured bass annotation using the new Bach Chorales Figured Bass Dataset},
  author={Ju, Yaolong and Margot, Sylvain and McKay, Cory and Dahn, L and Fujinaga, I},
  booktitle={Proceedings of the 21st International Society for Music Information Retrieval Conference},
  year={2020}
}

@inproceedings{liang2017automatic,
  title={Automatic Stylistic Composition of Bach Chorales with Deep LSTM.},
  author={Liang, Feynman T and Gotham, Mark and Johnson, Matthew and Shotton, Jamie},
  booktitle={ISMIR},
  volume={2017},
  pages={18th},
  year={2017}
}

@article{marinescu2019bach,
  title={Bach 2.0-generating classical music using recurrent neural networks},
  author={Marinescu, Alexandru-Ion},
  journal={Procedia Computer Science},
  volume={159},
  pages={117--124},
  year={2019},
  publisher={Elsevier}
}

@book{mattheson1731grosse,
  title={Grosse General-Bass-Schule},
  author={Mattheson, Johann},
  year={1731},
  publisher={Hamburg: Johann Christoph Kißner}
}

@article{mortensen1996unerringly,
  title={Unerringly tasteful'?: Harpsichord continuo in Corelli's op. 5 sonatas.},
  author={Mortensen, Lars Ulrik},
  journal={Early Music},
  volume={24},
  number={4},
  year={1996}
}

@misc{muffat1699regulae,
  author       = {Muffat, Georg},
  title        = {Regulae Concentuum Partiturae},
  year         = {1699},
  note         = {Wien: Musikarchiv Minoritenkonvent (Zentralbibliothek der {\"O}sterreichischen Minoritenprovinz), MS 1 B 7},
}

@article{niitsuma2011development,
  title={Development of a method for automatic basso continuo playing},
  author={Niitsuma, Masahiro and Matsubara, Masaki and Oono, Masaki and Saito, Hiroaki},
  journal={Information Processing \& Management},
  volume={47},
  number={3},
  pages={440--451},
  year={2011},
  publisher={Elsevier}
}

@article{scikit-learn,
  title={Scikit-learn: Machine Learning in {P}ython},
  author={Pedregosa, F. and Varoquaux, G. and Gramfort, A. and Michel, V.
          and Thirion, B. and Grisel, O. and Blondel, M. and Prettenhofer, P.
          and Weiss, R. and Dubourg, V. and Vanderplas, J. and Passos, A. and
          Cournapeau, D. and Brucher, M. and Perrot, M. and Duchesnay, E.},
  journal={Journal of Machine Learning Research},
  volume={12},
  pages={2825--2830},
  year={2011}
}

@misc{stefunko_2025_AIMC,
  author       = {Štefunko, Adam and
                  Chiruthapudi, Suhit and
                  Hajič jr., Jan and
                  Cancino-Chacón, Carlos Eduardo},
  title        = {Basso Continuo Goes Digital: Collecting and
                   Aligning a Symbolic Dataset of Continuo
                   Performance
                  },
  month        = sep,
  year         = 2025,
  publisher    = {Zenodo},
  doi          = {10.5281/zenodo.16946799},
  url          = {https://doi.org/10.5281/zenodo.16946799},
}

@misc{yao_2025_generality,
  author       = {Yao, Mingyang and
                  Chen, Ke},
  title        = {From Generality to Mastery: Composer-Style
                   Symbolic Music Generation via Large-Scale Pre-
                   training
                  },
  month        = sep,
  year         = 2025,
  publisher    = {Zenodo},
  doi          = {10.5281/zenodo.16948235},
  url          = {https://doi.org/10.5281/zenodo.16948235},
}

@ARTICLE{zhu_GPT,
  author={Zhu, Lin and Zhang, Wenjuan},
  journal={IEEE Access}, 
  title={Research on Polyphonic Music Generation Algorithm Based on GPT Large Model}, 
  year={2025},
  volume={13},
  number={},
  pages={129638-129647},
  doi={10.1109/ACCESS.2025.3588847}}

@inproceedings{wead2007computer,
  title={A computer-based implementation of basso continuo rules for figured bass realizations},
  author={Wead, Adam and Knopke, Ian},
  booktitle={ICMC},
  year={2007}
}





% \end{thebibliography}

\end{document}